\documentclass[12pt,a4paper,titlepage]{article}
\usepackage{t1enc}
\usepackage{amsmath}
\usepackage{apalike}
\usepackage{latexsym}
\usepackage{graphicx}
\usepackage{rotating}
\usepackage{dcolumn}
\usepackage{epsfig}
\usepackage{textcomp}
\usepackage{xypic,t1enc}

\usepackage{alltt}

\begin{document}
\newcommand{\diff}{\mathop{}\!\mathrm{d}}

\newcommand{\nothere}[1]{}
\newcommand{\noi}{\noindent}
\newcommand{\cip}{\perp\!\!\!\perp}

\newcommand{\cP}{\stackrel{P}{\rightarrow}}
\newcommand{\mbf}[1]{\mbox{\boldmath $#1$}}
\newcommand{\cond}{\, |\,}
\newcommand{\hO}[2]{{\cal O}_{#1}^{#2}}
\newcommand{\hF}[2]{{\cal F}_{#1}^{#2}}
\newcommand{\tl}[1]{\tilde{\lambda}_{#1}^T}
\newcommand{\la}[2]{\lambda_{#1}^T(Z^{#2})}
\newcommand{\I}[1]{1_{(#1)}}
\newcommand{\nrm}[1]{\| #1 \|_\infty}

\newcommand{\proof}[2]
{
\bigskip

\noindent {\bf Proof #1} $\,$ #2 \hfill $\Box$

\bigskip
}

\newcommand{\example}[1]
{
\bigskip

{ \sc Example.} #1 \hfill $\Box$

\bigskip
}

\newcommand{\remark}[1]
{
\bigskip

{\it  Remark.} #1 \hfill

\bigskip
}

\newtheorem{thm}{Theorem}
\newtheorem{lemma}{Lemma}
\newtheorem{prop}{Proposition}

\centerline{{\Large{\bf Instrumental variables estimation with competing risk data}}}
\vspace{2 cm}

{ \large
\begin{center}
TORBEN MARTINUSSEN\\
Section of Biostatistics\\
University of Copenhagen\\
\O ster Farimagsgade 5B, 1014 Copenhagen~K, Denmark\\vspace{4mm}
{tma@sund.ku.dk}\\[4pt]
{\sc And} \\
\vspace{4mm}
STIJN VANSTEELANDT\\[4pt]
Department of Applied Mathematics, Computer Sciences and Statistics,
Ghent University, Krijgslaan 281 (S9), 9000 Gent, Belgium, 
and Centre for Statistical Methodology, London School of Hygiene and Tropical Medicine, 
Keppel Street, London, WC1E 7HT, UK\\[2pt]
{stijn.vansteelandt@ugent.be}
\end{center}
}

\begin{abstract}
 {Time-to-event analyses are often plagued by both -- possibly unmeasured -- confounding and competing risks. To deal with the former, the use of instrumental variables for effect estimation is rapidly gaining ground. We show how to make use of such variables in competing risk analyses. In particular, we show 
how to infer the effect of an arbitrary exposure on cause-specific hazard functions under a semi-parametric model that imposes relatively weak restrictions on the observed data distribution. The proposed approach is flexible  accommodating
 exposures and instrumental variables of arbitrary type, and enables covariate adjustment. It
  makes use of closed-form estimators that can be recursively calculated, and is shown to perform well in simulation studies.
   We also demonstrates its use in
    an application on the effect of mammography screening on the risk of dying from breast cancer.}
 {Causal effect; Competing risk; Instrumental variable; Time-to-event; Unobserved confounding}
\end{abstract}

\section{Introduction}
\label{sec:Introduction}
In most observational studies unobserved confounding cannot be ruled out. This can make the results on exposure effects, as obtained via standard regression methods, questionable. 
Sometimes, however, it may be possible to estimate an exposure effect without (large sample) bias when an instrumental variable (IV) is available.
 This is a variable which is (a) associated with the exposure, (b) has no direct effect on the outcome other than through the exposure, and (c) whose association with the outcome is not confounded by unmeasured variables (see e.g. Hern\'an and Robins, 2006). Condition (a) is empirically verifiable, but conditions (b) and (c) are not. 

Instrumental variables estimation of exposure effects is well established for continuous outcomes
that obey linear models. One popular technique is 2SLS estimation, sometimes also referred to as the two-stage predictor substitution (2SPS) method (Cai et al., 2011). Here, the exposure variable is regressed on the instrument in the first stage, and then the outcome variable is regressed on the predicted exposure value in the second stage. The regression coefficient of the predicted exposure in the second stage is then interpreted as the exposure effect of interest. 

Recently there has been a focus on extending these methods to handle also right censored failure time data. 
Robins and Tsiatis (1991) initiated this work, but although they developed a general estimating equations-based method under structural accelerated failure time models, their proposal suffers from a lack of smoothness of the estimating equations due to the way how censoring is handled (Joffe et al., 2012). Tchetgen Tchetgen et al. (2015) developed an easy-to-use two-stage estimation approach under additive hazard models for event times, which works when the exposure obeys a location shift model; see Li, Fine and Brookhart (2015) for a related approach under a more restrictive model. Martinussen et al. (2017) generalised these methods by working under a less restrictive semiparametric structural cumulative failure time model, imposing no restrictions on distribution (or type) of instrument or exposure. Their proposal has the further advantage of enabling non-parametric estimation of a possibly time-varying exposure effect. Kjaersgaard and
 Parner (2015) suggested an alternative approach based on pseudo-observations. 
 Their 2SLS method requires a latent additive model for the target parameter which is not so attractive when focussing on a distribution function.
 
Motivated by an analysis of the HIP-study, which was designed to assess the potential effect of breast cancer screening,  we here aim at extending the methods of Martinussen et al. (2017) to handle competing risk data. 
The HIP-study comprised approximately 60000 women, who were randomised into two approximately equally sized groups. About 35\% of the women who were offered screening, refused to participate, resulting in a problem of non-compliance. We planned to correct for this using randomisation as an IV. In the first 10 years of follow-up, there were 4221 deaths, but only 340 were deemed due to breast cancer, making competing risks a major issue in these data. 

Richardson et al. (2017)  proposed a method that can deal with competing risk data also using an IV approach. Their suggestion  requires  the instrument as well as exposure variable to be binary variables. Essentially they generalise the standard IV Wald estimator for survival probabilities to estimate cumulative incidence probabilities. In this way they estimate the so-called complier treatment effect. The method we propose puts no restriction on the type of instrument nor on the exposure, and it can also incorporate covariates which is not possible using the Wald type estimator of Richardson et al. (2017). Our method is thus much more general. Zheng et al. (2017) suggest a method that directly models the subdistribution hazard, a quantity that is hard to interpret, see Andersen and Keiding (2012). Furthermore, Zheng et  al. (2017) requires a model for unobserved variables. Such models can never be checked and resulting estimators will be  purely model driven. 

The paper is structured as follows. In the next section we 
specify the model and outline the estimation procedure.
Section 3 contains large sample results.
In Section 4 we study by simulations the practical behavior of the proposed estimator and also analyse the HIP-data.
Section 5 contains some closing remarks and technical details are deferred to the Appendix.

\section{Model specification and estimation}

We let $\tilde T$ denote the time until one of the two competing events happens and let $\delta=1,2$ denote which of the two that takes place. 
Our aim is to assess the effect of an arbitrary exposure $X$ on the cause-specific hazard of each of these competing events, by making use of an instrumental variable $G$. This variable is such that, possibly conditional on measured covariates $L$, $G$ is associated with the exposure $X$, but is not associated with the event time $\tilde T$, nor the event type $\delta$, except because of a possible exposure effect. 
More formally, let $(\tilde T^x,\delta^x)$ denote the counterfactual event time and event type that would be observed for given subject if the exposure of that subject were set to $x$. We will make the consistency assumption that these coincide with the observed event time  $\tilde T$ and event type $\delta$ for those subjects who happen to have exposure level $X=x$. This notation enables us to be clear about our target of inference, which is the contrast between the counterfactual cause-specific hazard functions
\begin{equation}\label{estimand}
\lambda^j_{T^x}(t|X=x,G,L)-\lambda^j_{T^0}(t|X=x,G,L),\end{equation}
for $j=1,2$, where 
\begin{equation}
\label{cause_spec_haz}
\lambda_{T^x}^j(t|X,G,L)=\frac{ \frac{\diff}{\diff t}P(\tilde T^x\le t,\delta^x=j| X,G,L)}{P(\tilde T^x>t| X,G,L)},\quad j=1,2.
\end{equation}

Because $T^0$ and $\delta^0$ are unobserved for subjects with non-zero exposure, we will rely on the assumption that $G$ is an instrumental variable for the exposure effect (conditional on $L$), in the sense that $(\tilde{T}_0,\delta^0)$ is conditionally independent of $G$, given $L$ (Hern\'an and Robins, 2006). This assumption expresses that, if all subjects received zero exposure, both events would have cause-specific hazards (conditional on $L$) that would be same at all levels of $G$. This would be the case when, as in the causal Directed Acyclic Graph of Figure 1, $G$ shares no common causes with the event time and type, and does not influence those in the absence of exposure. As in other instrumental variables problems, these instrumental variables assumptions will not generally suffice to identify the contrast (\ref{estimand}) at all levels of $X,G$ and $L$. For that reason, as well as for reasons of parsimony, we will assume that  the following structural model holds
\begin{equation}
\label{StrucMod}
\lambda^j_{T^x}(t|X=x,G,L)-\lambda^j_{T^0}(t|X=x,G,L)=\beta_j(t)x, 
\end{equation}
for all $t>0$ and for $j=1,2$, with $\beta_j(t)$ an unknown, locally integrable function. Our aim is then to estimate $B_j(t)=\int_0^t\beta_j(s)\, ds$ for all $t>0$, $j=1,2.$
As in Martinussen et al. (2017), it can be shown that this model is satisfied when the causal Directed Acyclic Graph of Figure 1 holds and, moreover, 
$$
\lambda^j(t|X,G,L,U)=\lambda_{j}(t)+\beta_{j}(t)X+\psi_j(t,U,L)
$$
with the function $\psi_{j}(.),j=1,2$ left unspecified and the functions $\lambda_{j}(.)$ and $\beta_j(t)$ unknown.
Note however that model (\ref{StrucMod}) is less restrictive; e.g. it makes no assumptions about the dependence of the event time on the unmeasured confounders $U$.
Our proposal may be extended  for instance to allow for interactions between $X$ and $L$ but we will focus on the simple setting to keep expressions more transparent.

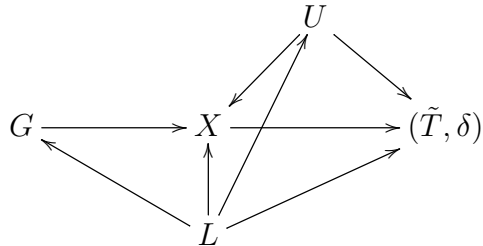
\begin{figure}[h!]
$$
\xymatrix{& &&   U \ar[dl]\ar[dr] &\\
G\ar[rr]&  &X  \ar[rr] & &(\tilde T,\delta)\\
& & \ar[ull] L \ar[u] \ar[urr] \ar[uur] & &\\
}
 $$
  \caption{Causal Directed Acyclic Graph. $G$ is is the instrument, $X$ the exposure variable and $\tilde T$ the time-to-event and $\delta$ indicates which of the two competing events that is taking place. The potential unmeasured confounders are denoted by $U$, and the observed confounders by $L$.}
\end{figure}
 
Throughout, we will allow for the event time $\tilde T$ to be subject to right-censoring. In that case, we only observe whether or not $\tilde T$ exceeds a random 
censoring time $C$, i.e. we observe $D=I(\tilde T\leq C)$, along with the first time either failure or censoring occurs, i.e. we also observe $T=\min(\tilde T,C)$ as well as $\delta$ if $D=1$. Let $\delta=0$ if $D=0$. Define also the observed counting processes 
$N^j(t)=I(T\leq t,D=1,\delta=j)$, $j=1,2$, and  the at risk indicator $R(t)=I(t\leq T)$.
We  assume that the censoring time satisfies the following condition
\begin{equation}
\tag{C}
\mbox{$\tilde T\cip C|X,G,L$  and $P(C>t|X,G,L)=P(C>t|L)$}
\end{equation}
The above condition on the censoring distribution can be relaxed to  $P(C>t|X,G,L,U)=P(C>t|L,U)$ for some variable $U\cip G|L$.

The following Proposition lays the basis of the estimation procedure for $B_j(t)$, which we will describe next.
\begin{prop}
Assume the structural model \eqref{StrucMod} with the assumption that $G$ is an instrumental variable, conditional on $L$,  and further that the censoring time satisfies condition (\cal{C}).
Then  
\begin{equation}
\label{Esteq}
E\left[\left\{G-E(G|L)\right\}e^{B_1(t)X+B_2(t)X}R(t)\left\{dN^j(t)-dB_j(t)X\right\}\right]=0,
\end{equation}
for each $t$, $j=1,2$.
\end{prop}
\medskip

\noindent
{\it Proof.}
 By the independent censoring assumption (\cal{C})  and
 $$
 \frac{P(\tilde T^0>t|X,G,L)}{P(\tilde T>t|X,G,L)}=e^{B_1(t)X+B_2(t)X}
 $$
 it follows,
for $j=1,2$,
 that 
\begin{align*}
&E\left[\left\{G-E(G|L)\right\}e^{B_1(t)X+B_2(t)X}R(t)\left\{dN^j(t)-dB_j(t)X\right\}\right]\\
&=E\left[\left\{G-E(G|L)\right\}e^{B_1(t)X+B_2(t)X}I(C>t)I(\tilde T>t)\lambda^j_{\tilde{T}^0}(t|X,G,L)dt\right]\\
&=E\left[P(C>t|L)\left\{G-E(G|L)\right\}P(\tilde T^0>t|X,G,L)\lambda^j_{\tilde{T}^0}(t|X,G,L)dt\right]\\
&=E\left[P(C>t|L)\left\{G-E(G|L)\right\}\frac{d}{dt}P(\tilde T^0\leq t,\delta^0=j|X,G,L)dt\right]\\
&=\frac{d}{dt}E\left[P(C>t|L)\left\{G-E(G|L)\right\}P(\tilde T^0\leq t,\delta^0=j|G,L)dt\right]\\
&=0
\end{align*}
because, for any function $g_t(L)$, we have 
$$
E\left[g_t(L)\left\{G-E(G|L)\right\}P(\tilde T^0\leq t,\delta^0=j|G,L)dt\right]=0
$$ 
since $G\cip (\tilde{T}^0,\delta^0)|L$. This completes the proof. $\Box$
\bigskip

The above proposition gives two unbiased estimating functions for each time $t$, on the basis of which we can construct a consistent estimator of $B_j(t)$. In particular, let $(T_i,D_i,\delta_i,L_i,G_i,X_i)$, $i=1,\ldots, n,$ denote $n$ independent identically distributed replicates under the structural  model \eqref{StrucMod} together with the instrumental variables assumptions. 
Suppose that the counting processes $N^j_i(t)=I(T_i\leq t,D_i=1,\delta_i=j)$, $i=1,\ldots, n$, $j=1,2$,  are observed in the time interval $[0,\tau ]$, where $\tau$ is some finite time point. Solving equation \eqref{Esteq}, with population expectations substituted by sample analogs leads to
 the recursive estimator $\hat B_j(t)$ defined by
\begin{equation}
\label{Est}
\hat B_j(t,\hat \theta)=\int_0^t \frac{\sum_iG^c_i(\hat\theta)e^{\{\hat B_1(s-)+\hat B_2(s-)\}X_i}dN^j_i(s)}{\sum_iG^c_i(\hat\theta)R_i(s)e^{ \{\hat B_1(s-)+\hat B_2(s-)\}X_i}X_i},
\end{equation}
where $G^c_i(\theta)=G_i-E(G_i|L_i;\theta)$, with $E(G_i|L_i;\theta)$ a parametric model for $E(G_i|L_i)$ and $\hat\theta$ a consistent estimator of $\theta$ (e.g., a maximum likelihood estimator). Note that $\hat B_j(t,\hat \theta)$ is step function that is well defined by setting $\hat B_j(0,\hat \theta)=0$, $j=1,2$. Furthermore, note that  -- unlike many other IV-estimators -- the estimator (\ref{Est}) can be evaluated for discrete as well as continuous exposures and instruments. It does not require distributional assumptions for the exposure and does not make assumptions as to how measured covariates relate to the event time.

\section{Large sample properties}

The following proposition, whose proof is sketched  in the Appendix, shows that the estimators $\hat B_j(t)$, $j=1,2$,  are uniformly consistent. It moreover gives the asymptotic distribution of the two estimators.

\begin{prop}
Under model \eqref{StrucMod} with the assumption (C) and the assumption that $G$ is an instrumental variable, conditional on $L$, and given the technical conditions listed  in the Appendix, the  IV estimators  $\hat B_{j}(t)$, $j=1,2$, are  uniformly consistent. Furthermore,
 $
 W^j_n(t)=n^{1/2}\{\hat B_{j}(t,\hat \theta)-B_j(t)\}
 $
 converges in distribution to a zero-mean Gaussian process with variance $\Sigma_j(t)$. A uniformly consistent estimator $\hat\Sigma_j(t)$ of $\Sigma_j(t)$ is given below.
\end{prop}
\medskip

Let $\epsilon_i^{B_j}(t,\theta),i=1,...,n$ be the iid zero-mean processes given by expression \eqref{Eps} in the Appendix.
From the proof in the Appendix, it then follows that
$W^j_n(t)$ is asymptotically equivalent to
$n^{-1/2}\sum_{i=1}^n\epsilon_i^{B_j}(t,\theta)$.
 The variance $\Sigma_j(t)$ of the limit distribution can thus be consistently estimated by
\begin{equation}
\label{Variance}
\hat \Sigma_j(t)=n^{-1}\sum_{i=1}^n\{\hat \epsilon_i^{B_j}(t,\hat\theta)\}^2,
\end{equation}
where $\hat \epsilon_i^{B_j}(t,\hat \theta)$ is obtained from $\epsilon_i^{B_j}(t,\theta) $ by replacing unknown quantities with their empirical
counterparts.
These results
can be used to construct a pointwise confidence band.

\section{Numerical results}
 In this section we investigate the practical behavior of the proposed estimator. In Section 4.1 we study the small sample performance using simulations, and in Section 4.2 we give a worked application using the HIP-data on the potential effect of breast cancer screening on death due to breast cancer.
\subsection{Simulation study}
To investigate the properties of our proposed methods with practical sample sizes we conducted a simulation experiment, whereby we generated data under the causal Directed Acyclic Graph of Figure 1.
 We took $G$ to be binary with $P(G=1)=0.5$, and generated $X$ and $U$, given $G$, from a normal distribution with 
$E(X|G=g)=0.5+\gamma_Gg$, $E(U|G=g)=1.5$ and with variance-covariance matrix so that $Var(X|G)=Var(U|G)=0.25$, and $Cov(X,U|G)=-1/6$.
The parameter  $\gamma_G$  determines the size of the correlation between exposure and the instrumental variable. Specifically we looked at correlation $\rho$ equal to 0.3 and 0.5.
The two cause specific hazards were given as
\begin{align*}
\lambda^1(t|X,G,U)&=0.1+0.1U\\
\lambda^2(t|X,G,U)&=0.1+0.2X+0.1U
\end{align*}
so $\tilde T$ is generated according to the hazard 
$$\lambda_{\tilde T}(t|X,G,U)=0.2+0.2X+0.2U$$ and failure of type 1 happens with probability $$\frac{\lambda^1(t|X,G,U)}{\lambda^1(t|X,G,U)+\lambda^2(t|X,G,U)},$$ and likewise with failure of type 2. It is easily seen that model (2) holds under this model.
Twenty percent were potentially censored according to a uniform distribution on (0,3.5), and the rest were censored at $t=3.5$, corresponding to the study being closed at this time point, leading to an overall censoring rate of around 17.
\nothere{
Under this model, as seen in the Section 2.2,  \eqref{Param} 
holds with $B_X(t)=\int_0^t\beta_X(s)\; ds=0.1t$. 
Under this model it further holds that 
$$
E\left\{d\tilde N(t)|T\geq t, X,G\right\}=\tilde\beta_0(t)+\tilde\beta_G(t)G+\tilde\beta_X(t)X,
$$
with $\tilde\beta_X(t)=0$ so the naive Aalen estimator (using $X$ and $G$ as covariates) is biased.
We calculated the estimator given in (\ref{Est}) with $\hat \theta=\overline{G}$, along with the estimator $\hat \beta_X$ given in (9) where we took $\tau=3$.
}
For this scenario, we considered sample size 1600 when $\rho=0.3$, and sample size 1000 when $\rho=0.5$.
 Simulation results  are given in Table 1 based on 2000 runs for each configuration, where we report 
 average biases at time points $t=0.5, 1.5, 2.5$ for $\hat B_{j}(t)$. We also report the empirical standard errors as well as estimated standard errors based on formula \eqref{Variance}
  along with coverage probability of 95\% pointwise confidence intervals CP($\hat B_{j}(t)$). Biases from the naive Aalen estimator, denoted as $\tilde B_{j}(t)$ in the table, running Aalen's additive hazards model  (see Martinussen and Scheike, 2006, Ch. 5) for the two cause specific hazards using $X$ and $G$ as covariates are also given.

\vspace{0.5cm}
\centerline{Table 1 about here}
\vspace{0.5cm}

\begin{table}[h!]
\caption{Continuous exposure and binary instrument.
The two causes are given by $j=1,2$.
Bias of $\hat B_{j}(t)$, average estimated standard error, sd($\hat B_{j}(t)$), empirical standard error, see($\hat B_{j}(t)$), and coverage probability of 95\% pointwise confidence intervals CP($\hat B_{j}(t)$) based on the instrumental variables estimator, in function of sample size $n$ and at different strengths $\rho$ (correlation) of the instrumental variable. Bias of $\tilde B_{j}(t)$ is the bias of the naive Aalen estimator.
 \bigskip}

{\small
\begin{center}
\begin{tabular}{|l|cccc|cccc|}
\hline
 &&\multicolumn{3}{c|}{ $\rho=0.3$}&&\multicolumn{3}{c|}{ $\rho=0.5$} \\
&n & $t=0.5$ & $t=1.5$ & $t=2.5$ & n& $t=0.5$ & $t=1.5$ & $t=2.5$\\

\hline
Bias $\hat B_{1}(t)$ &1600&0.001   &   0.007     &  -0.007&1000&       -0.000&-0.001  &  - 0.004      \\
sd ($\hat B_{1}(t)$)&&0.061 &0.132& 0.233& &0.042&0.089 &0.149\\
see ($\hat B_{1}(t)$)&&0.061&0.130& 0.231& &0.042&0.089 &0.149\\
95\% CP($\hat B_{1}(t)$)&&95.4&     95.3&      96.6   &&   95.0 &95.5&     95.7\\
Bias $\hat B_{2}(t)$ &&0.002   &   -0.001     &  -0.010& &       0.000&-0.000   &  - 0.005      \\
sd ($\hat B_{2}(t)$)&&0.077 &0.165& 0.300& &0.053&0.113 &0.198\\
see ($\hat B_{2}(t)$)&&0.076 &0.165& 0.296& &0.054&0.117 &0.199\\
95\% CP($\hat B_{2}(t)$)&&94.8&     95.6&      96.0   &&   95.2 &96.1&     96.5\\
Bias $\tilde B_{1}(t)$ &&-0.033   &   -0.099    &  -0.167& &      -0.034&-0.101   &  - 0.168      \\
Bias $\tilde B_{2}(t)$ &&-0.034   &   -0.102     &  -0.170& &      -0.035&-0.102   &  - 0.168      \\
\hline
\end{tabular}
\end{center}
}
\end{table}
\bigskip

\noindent
It is seen from Table 1 that the suggested estimators are unbiased and also that the estimated standard errors estimate well the variability resulting in satisfactory coverage probabilities. As expected, the naive estimators $\tilde B_{j}(t)$, $j=1,2$, are biased. 

We also considered a simulation scenario where we took both the exposure variable and the instrument to be continuous variables. To our knowledge there are no other available methods to handle such a situation. Specifically, we generated data as in the first simulation study with the difference that
$G$ was now standard normal, and $X$ and $U$ were generated, given $G$, from a normal distribution with 
$E(X|G=g)=1.5+\gamma_Gg$, $E(U|G=g)=1.5$ and with variance-covariance matrix so that $Var(X|G)=Var(U|G)=0.25$, and $Cov(X,U|G)=-1/6$.
The parameter  $\gamma_G$  determines the size of the correlation between exposure and the instrumental variable. Specifically we looked at correlation $\rho$ equal to 0.3 and 0.5.  We did 2000 runs for each configuration.

\vspace{0.5cm}
\centerline{Table 2 about here}
\vspace{0.5cm}

\begin{table}[h!]
\caption{Continuous exposure and instrument.
The two causes are given by $j=1,2$.
Bias of $\hat B_{j}(t)$, average estimated standard error, sd($\hat B_{j}(t)$), empirical standard error, see($\hat B_{j}(t)$), and coverage probability of 95\% pointwise confidence intervals CP($\hat B_{j}(t)$) based on the instrumental variables estimator, in function of sample size $n$ and at different strengths $\rho$ (correlation) of the instrumental variable. Bias of $\tilde B_{j}(t)$ is the bias of the naive Aalen estimator.
 \bigskip}

{\small
\begin{center}
\begin{tabular}{|l|cccc|cccc|}
\hline
 &&\multicolumn{3}{c|}{ $\rho=0.3$}&&\multicolumn{3}{c|}{ $\rho=0.5$} \\
&n & $t=0.5$ & $t=1.5$ & $t=2.5$ & n& $t=0.5$ & $t=1.5$ & $t=2.5$\\

\hline
Bias $\hat B_{1}(t)$ &1600&0.001   &   -0.001     &  -0.008&1000&       -0.000&-0.001  &  - 0.005      \\
sd ($\hat B_{1}(t)$)&&0.052 &0.112& 0.195& &0.038&0.081 &0.132\\
see ($\hat B_{1}(t)$)&&0.053&0.113& 0.223& &0.038&0.080 &0.148\\
95\% CP($\hat B_{1}(t)$)&&95.6&     96.1&      97.6   &&   95.5 &95.4&     96.8\\
Bias $\hat B_{2}(t)$ &&-0.002   &   -0.004     &  -0.002& &       0.000&-0.002  &  - 0.003      \\
sd ($\hat B_{2}(t)$)&&0.072 &0.147& 0.251& &0.049&0.104 &0.179\\
see ($\hat B_{2}(t)$)&&0.070 &0.148& 0.284& &0.049&0.104 &0.186\\
95\% CP($\hat B_{2}(t)$)&&95.0&     95.9&      97.6   &&   95.4 &95.1&     96.4\\
Bias $\tilde B_{1}(t)$ &&-0.027   &   -0.080   &  -0.134& &      -0.026&-0.081   &  - 0.132      \\
Bias $\tilde B_{2}(t)$ &&-0.032   &   -0.100    &  -0.1765& &      -0.033&-0.099   &  - 0.166      \\
\hline
\end{tabular}
\end{center}
}
\end{table}
\bigskip

\noindent
Similar conclusions are obtained from Table 2,
the suggested estimators are unbiased and also that the estimated standard errors are reasonably close to the empirical standard deviations although being a little too large at the later time point, $t=2.5$. Again, the naive estimators $\tilde B_{j}(t)$, $j=1,2$, are biased.

\subsection{Application to the HIP trial on effectiveness of screening  on breast cancer mortality}
The Health Insurance Plan (HIP) of Greater New York was a randomized trial of breast cancer screening that began in 1963. 
%
About 60000 women aged 40-60 were randomized into two approximately equally sized groups. Study women were offered the screening examinations consisting of clinical examination, 
and a mammography. Further three annual examinations were offered in this group. Control women continued to receive their usual medical care. 
There were 30565 women in the control group and 30130 in the screening group of which 9984 (35\%) 
 refused to participate  (non-compliers).
There were  large differences between the study women who participated and those who refused (Shapiro, 1977) and therefore the  results from the `as treated' analysis may be doubtful  due to potential unobserved confounding.

 \begin{figure}
\begin{center}
\includegraphics[width=16cm, height=8cm]{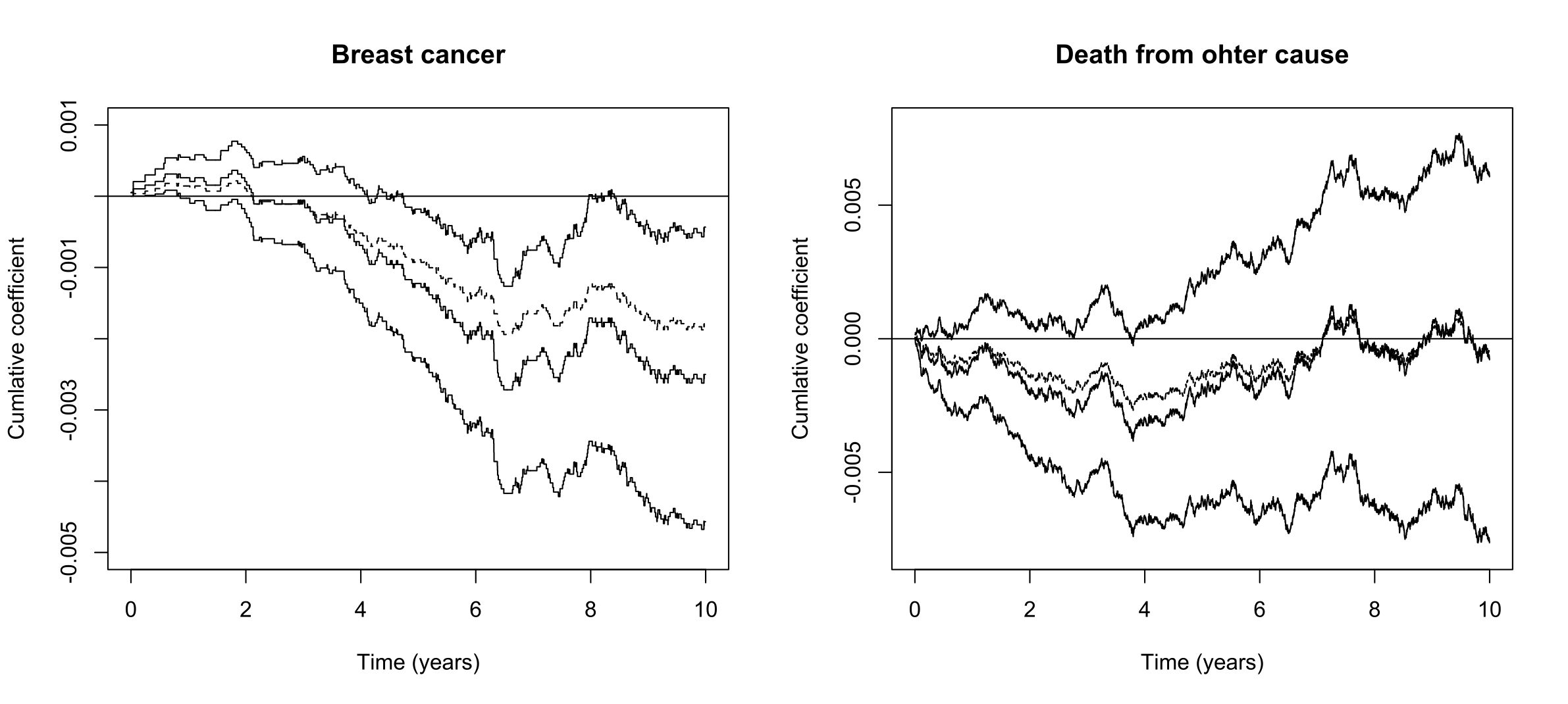}
 \caption{HIP-study. 
 Estimated causal effect of screening, $\hat B(t)$   along with 95\% pointwise confidence bands and the intention to treat estimate (broken curve). Curves are given for the two competing events.
 }
 \end{center}
 \end{figure}
 
 \noindent
 We applied the estimator given by \eqref{Est} to these data focussing on the first 10 years of follow-up. This estimator is shown in Figure 2 along with 95\% pointwise confidence intervals. Left panel gives results for breast cancer and right panel for other causes. The intention to treat estimator is also shown (dotted curves). It is seen from Figure 2 that breast cancer screening appears to lower the risk of dying from breast cancer while there is no evidence of an effect of screening on the risk of dying from other causes.
 The impact of the screening on the risk of dying from breast cancer seems to be slightly more pronounced than what is indicated by the intention to treat estimator. 
As the specific value of $B_1(t)$ may be hard to interpret we suggest also to report relative risks.
We have 
$$
\lambda^j_{T^0}(t|X=1,G,L)=\lambda^j_{T^1}(t|X=1,G,L)-\beta_j(t)
$$
and since, for these data, it seems reasonable that $\beta_2(t)=0$, we then have that for women who received screening on the active arm the relative risk of dying from cause 1 by time $t$ without versus with screening, 
\begin{align}
\label{RR}
\mbox{RR}(t)\equiv \frac{P(T^0\le t,\delta^0=1| X=1,G=1)}{P(T^1\le t,\delta^1=1| X=1,G=1)}
\end{align}
can be expressed as
\begin{align*}
&\frac{\int_0^te^{-\Lambda^1_{T^0}(s|X=1,G=1)-\Lambda^2_{T^0}(s|X=1,G=1)}d\Lambda^1_{T^0}(s|X=1,G=1)ds}{P(T\le t,\delta=1| X=1,G=1)} \\
=&\frac{\int_0^te^{-\Lambda^1_{T}(s|X=1,G=1)-\Lambda^2_{T}(s|X=1,G=1)+B_1(s)}\{d\Lambda^1_T(s|X=1,G=1)-dB_1(s)\}}{
\int_0^t P(T>s,\delta=1| X=1,G=1)d\Lambda^1_T(s|X=1,G=1)} 
\end{align*}
 which can be estimated using the proposed estimator, and by performing an all cause mortality analysis and a cause 1 (breast cancer) specific analysis conditioning on $X=G=1$. Such analyses results in  estimators $\hat \Lambda^j_{T}(t|X=1,G=1)$
 and $\hat P(T>t,\delta=1| X=1,G=1)$.
 Furthermore, we can evaluate  the variability of these components, and can then also combine these to estimate the variability of  
 $\hat{ \mbox{RR}}(t)$.
 Figure 3 displays $\hat{ \mbox{RR}}(t)$ along with 95\% pointwise confidence bands. It is seen that for women who received screening on the active arm the risk of dying from breast cancer within 5 (10) years would have been approximately twice (1.5 times) as large
 had they not received screening.


  \begin{figure}
\begin{center}
\includegraphics[width=12cm, height=9cm]{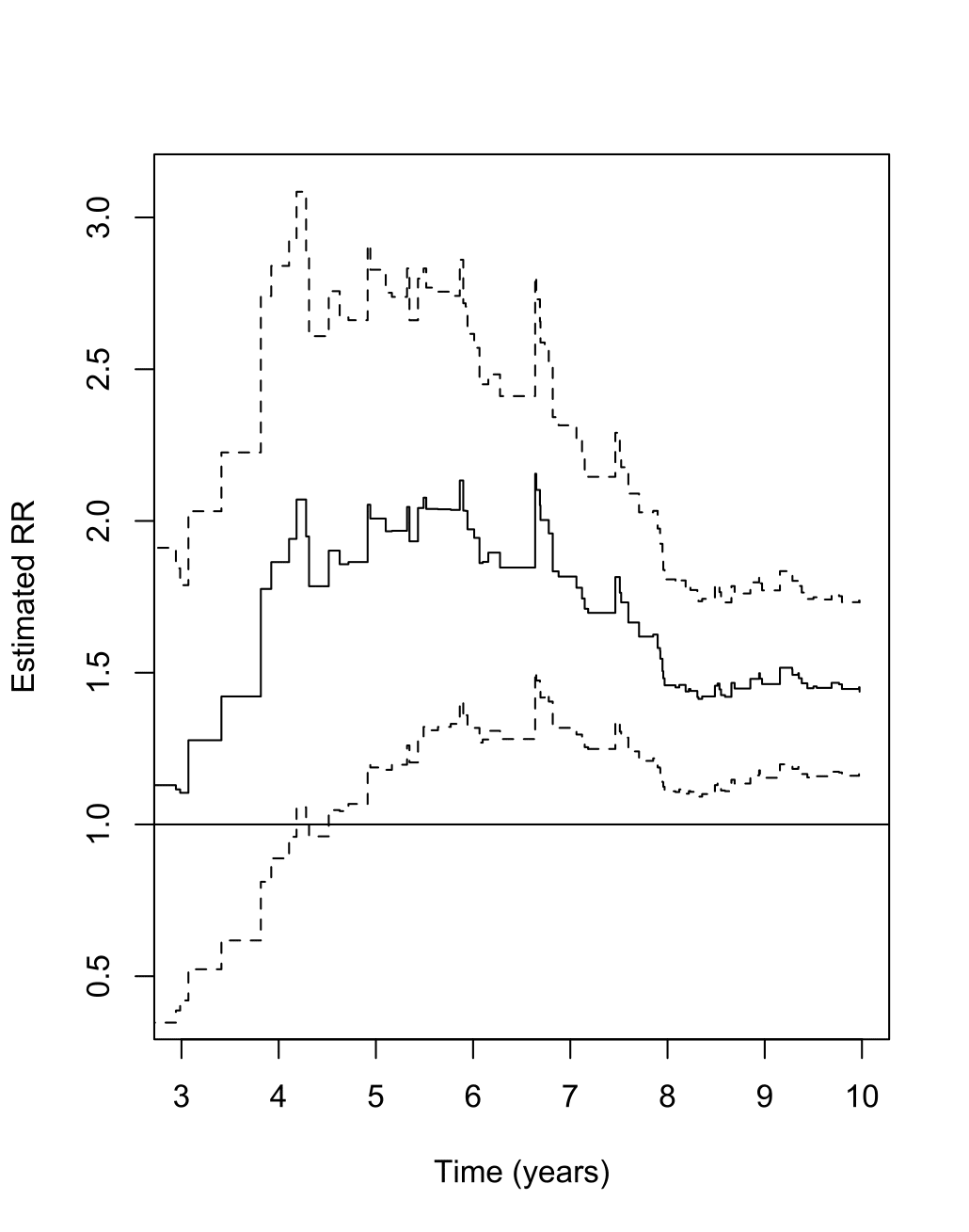}
 \caption{HIP-study. 
 Estimated relative risk function (breast cancer death), see display \eqref{RR}, along with 95\% pointwise confidence bands (dotted curves).
 }
 \end{center}
 \end{figure}

\section{Concluding remarks}

In this paper, we have proposed an approach to 
estimate causal effects in a competing risk setting where there may be unobserved confounding.
The proposal is based on the availability of an instrumental variable.
It can accommodate adjustment for baseline covariates, which is sometimes needed to make the instrumental variables assumptions more plausible. Unlike available instrumental variable methods, it makes no restriction on the type, nor the distribution of exposure or instrument. 
We can in particular deal with a situation where the exposure is continuous and the instrument is categorical, or where both are continuous.
Dichotomisation of the exposure, which is sometimes considered by simpler proposals, is no valid remedy in such cases as it entails a violation of the exclusion restriction.

One further strength of the approach is that it naturally adjusts for censoring whenever censoring is independent of the event time conditional on exposure, instrument and confounders, as well as 
when censoring is independent of the exposure and instrument conditional on the confounders.
 Although the latter assumption could fail, it can 
 be remedied by applying inverse probability of censoring weighting; that is, 
 by redefining
 $
 e^{B_1(t)X+B_2(t)X}R(t)
 $
 in \eqref{Esteq} to 
  \[e^{B_1(t)X+B_2(t)X}R(t)\frac{P(C\geq t|\mathcal{F}^N_{t}\vee L)}{P(C\geq t|\mathcal{F}^N_{t}\vee X\vee G\vee L)},\]
where $\mathcal{F}^N_{t}$ denotes the history spanned by the counting processes.
Using this modification requires postulating two models for the cause-specific hazard of censoring: one conditional on $X,G$ and $L$, and one conditional on $L$ only. However, only the former model must be correctly specified to maintain a consistent estimator.

The results in this paper may be  extended to more general models than (\ref{StrucMod}), to also handle interactions with the confounder $L$. For instance, suppose that instead of (\ref{StrucMod}) the following model holds, for $j=1,2$,
\begin{equation}
\label{Model2}
\lambda^j_{T^x}(t|X=s,G,L)-\lambda^j_{T^0}(t|X=x,G,L)=\beta_j(t)x+\beta^T_{jXL}(t)xl
\end{equation}
where $\beta_{jXL}(t)$ is of dimension corresponding to $L$, say $p$. 
Let $B_{\ast}(t)$ denote the 
integral from 0 to $t$ of the corresponding $\beta_{\ast}(t)$, and define
 $B(t)=\{B_1(t),B_2(t),B^T_{1XL}(t),B^T_{2XL}(t)\}$. Let $dN(t)=\{dN_1^1(t),dN_1^2(t),\ldots ,dN_n^1(t),dN_n^2(t)\}^T$. We can the write the estimator of $B(t)$ as
 $$
 \hat{B}(t,\theta)=\int_0^t\left [\{J_1(s,\theta)J_2(s,\theta)\}^T\{J_1(s,\theta)J_2(s,\theta)\}\right ]^{-1}[\{J_1(s,\theta)J_2(s,\theta)\}^TJ_1(s,\theta)dN(s),
 $$
 where $J_1(t,\theta)$ is $1\times 2n$-matrix with $i$th row 
 $$
 \left\{G_i-E(G_i|L_i)\right\}\exp{\left \{\sum_{j=1}^2 \hat{B}_j(t-,\theta)X_i+ \hat{B}^T_{jXL}(t-,\theta)X_iL_i\right\}}R_i(t),
 $$
 and $J_2(t)$ is $2n\times (2+2p)$-matrix consisting of $n$ blocks of size $2\times (2+2p)$, where the $i$th block is
$$
\begin{bmatrix}
    X_i      & 0  & X_iL_i^T & 0 \\
  0     & X_i & 0& X_iL_i^T
\end{bmatrix}.
 $$

\section*{Acknowledgement}
Torben Martinussen's work is part of the Dynamical Systems Interdisciplinary Network, University of Copenhagen.
Stijn Vansteelandt was supported by IAP research network grant nr. P07/05 from the Belgian government (Belgian Science Policy).	


\section*{Appendix: Large sample properties}

\noindent
{\bf Consistency}
\bigskip
\medskip

Let $\mu(L;\theta)=E(G|L;\theta)$ be the conditional mean of the instrument given observed confounders $L$, which is function of an unknown finite-dimensional parameter $\theta$. In the case of no observed confounders $\mu(\theta)=\theta=E(G)$ and $\hat \theta=\overline{G}$. We assume that $n^{1/2}(\hat\theta-\theta)=n^{-1/2}\sum_i\epsilon^{\theta}_i+o_p(1)$, where the $\epsilon^{\theta}_i$'s are zero-mean iid variables.

We write $\nrm{g} = \sup_{t \in [0,\tau]} |g(t)|$ 
Let $B^\circ_j(t)$ denote the true value of $B_j(t)$, and let $M_j^\circ = \nrm{B_j^\circ}<\infty$.
Conditions:
\begin{itemize}
\item[(i)] We assume that $X$ and $G$ are bounded.
\item[(ii)] Define $a(s,h) = E[R(s)XG^c e^{h X}]$. We assume that
there exist $M > M_j^\circ$, $j=1,2$, and $\nu>0$ such that
$\inf_{s \in [0,\tau], h \in[-M,M]} a(s,h) \geq 1.01 \nu$. 
\end{itemize}
The quantities $M_j^\circ$ and $M$ do not necessarily need to be known.
Under these assumptions we may
modify the  arguments
 given in Martinussen et al. (2017) to also cover the competing risk situation described here. Hence, consistency can be inferred similarly.
\bigskip
\medskip

\noindent
{\bf Asymptotic normality}
\bigskip

\noindent
Let $N(t)$ be the $n\times 2$ matrix with $i$th row $\{N_i^1(t),N_i^2(t)\}$ and $X=(X_1,\ldots , X_n)^T$. For known $\theta$ we can write 
$$\hat B(t,\theta)=\{\hat B_1(t,\theta),\hat B_2(t,\theta)\}=\int_0^tH_{\theta}\{s,\hat B_1(s-,\theta)+\hat B_2(s-,\theta)\}dN(s),
$$
where the $k$th element of the $n$-vector $H_{\theta}\{t,a\}$ 
is 
$$R_k(t)G_k^c(\theta)e^{aX_k}/\sum_{i=1}^nG_i^c(\theta)R_i(t)e^{aX_i}X_i.
$$
with $a=\hat B_1(t-,\theta)+\hat B_2(t-,\theta)$.
Let $W_n(t,\theta)=n^{1/2}\{\hat B(t,\theta)- B(t)\}$, $b=(1,1)^T$, and let $\dot{H}$ denote the derivative of $H$ with respect to its second argument. It is then easy to see that 
\begin{align*}
W_n(t,\theta)=&n^{1/2}\int_0^tH\{s, B_1(s-)+B_2(s-)\}\left [dN(s)-XdB(s)\right ]\\
&+\int_0^tV(s-,\theta)\{1+o_p(1)
\}b\dot{H}\{s,B_1(s-)+B_2(s-)\}dN(s) 
\end{align*}
which is a  Volterra-equation, see Andersen et al. (1993), p. 91. The solution to this equation is given by
$$
W_n(t,\theta)=\int_0^tn^{1/2}H\{s,B_1(s-)+B_2(s-)\}\left [dN(s)-XdB_X(s)\right ]{\cal F}(s,t)+o_p(1),
$$
where 
$$
{\cal F}(s,t)=\prod_{(s,t]}\left \{I+b\dot{H}(\cdot,B_1(\cdot )+B_2(\cdot ))dN(\cdot)\right \}
$$
with the latter being a  product integral that converges in probability to some limit. This leads to the iid-representation
$$W_n(t,\theta)=n^{-1/2}\sum_{i=1}^n\epsilon^B_i(t)$$ with the $\epsilon^B_i(t)$'s being
zero-mean iid terms. Specifically
$$
\epsilon^B_i(t)=\int_0^tn^{1/2}\{H\{s,B_1(s-)+B_2(s-)\}\}_i\left [dN(s)-XdB_X(s)\right ]_i{\cal F}(s,t)
$$
with $a_i$ being the $i$th element of the vector $a$, and $[A]_i$ being the $i$th row of the matrix $A$.
This
together with
\begin{align*}
n^{1/2}\{\hat B(t,\hat\theta)-B(t)\}&=n^{1/2}\{\hat
B(t,\theta)-B(t)\}+n^{1/2}\{\hat B(t,\hat\theta)-\hat B(t,\theta)\}\\
&=n^{1/2}\{\hat B(t,\theta)-B(t)\}+D_{\theta}(\hat
B(t,\theta)^T)_{|\hat\theta}n^{1/2}(\hat\theta-\theta)+o_p(1),
\end{align*}
where $D_{\theta}\{\hat
B(t,\theta)^T\}$ is the first order derivative of $\hat
B(t,\theta)^T$ w.r.t. $\theta$ 
gives an iid-decomposition of $n^{1/2}\{\hat B(t,\hat\theta)-B(t)\}$:
$$
n^{1/2}\{\hat
B(t,\hat\theta)-B(t)\}=n^{-1/2}\sum_{i=1}^n\epsilon_i^B(t,\theta)+o_p(1),
$$
where 
\begin{equation}
\label{Eps}
\epsilon_i^B(t,\theta)=\epsilon^B_i(t)+D_{\theta}(\hat
B(t,\theta)^T)_{|\theta}\epsilon^{\theta}_i.
\end{equation}
It thus follows that
$$
n^{1/2}\{\hat B(t,\hat\theta)-B(t)\}
$$
converges to a zero-mean Gaussian process with a variance that is
consistently estimated by
$$
n^{-1}\sum_{i=1}^n\hat\epsilon_i^B(t,\hat\theta)^2.
$$
The derivative $D_{\theta}(\hat
B(t,\theta))_{|\hat\theta}$ can be calculated recursively as $\hat B(t,\hat\theta)$ is constant between the observed death times. Denote the jump times by $\tau_1,\ldots ,\tau_m$. Hence
$$
\hat B(\tau_j,\theta)=\hat B(\tau_{j-1},\theta)+d\hat B(\tau_j,\theta)
$$
which then also holds for the derivative. Since $\hat B(0,\theta)=0$ and the derivative of the increment in the first jump time,  $d\hat B(\tau_1,\theta)$, is easily calculated we then have a recursive way of calculating the derivatives of $\hat B(\cdot,\theta)$.
We now  argue that the process $W_n^j(t,\theta)$, $j=1,2$, converges in distribution  as a process using arguments similar to what is done in  Lin et al.  (2000.  p. 726).
It is seen from (\ref{StrucMod})
 that  $B_j(t)$ can be written as a difference of two monotone functions.
 Let $\tilde H_{ijk}(s)$ be the limit in probability of ${\cal F}_{kj}(s,t)H_i(s,B_X(s-))$. 
Now, split 	$\tilde H_{ijk}(s)$	 into its positive and negative parts, $\tilde H_{ijk}^+(s)$ and $\tilde H_{ijk}^-(s)$, 
and similarly with $X_i$, $X_i^+$ and $X_i^-$. 
Then $\int_0^t\tilde H_{ijk}(s)[ dN_i(s)-X_idB_X(s) ]_{ik}$ can be written as  a difference of two monotone functions, and then we follow the arguments of Lin et al. (2000) (or use example 2.11.16 of van der Vaart and Wellner, 1996).
Convergence in distribution for the process  $W_n^j(t,\hat \theta)$ also holds using the above Taylor expansion.
It thus follows that
$$
n^{1/2}\{\hat B(t,\hat\theta)-B(t)\}
$$
converges to a zero-mean Gaussian process.

\section*{References}

\noindent	Andersen, P. K. and Keiding, N. (2012). 
Interpretability and importance of functionals in competing risks and multistate models. \textit{Statistics in Medicine}
{\bf 31}, 1074-1088.
 \medskip
 
\noindent	Angrist, J. and Imbens, G. (1991). 
Sources of identifying information in evaluation models. 
Technical Working Paper 117, National Bureau of Economic Research, Cambridge, MA.
 \medskip
 
 \noindent Cai, B., Small, D. S. and Ten Have, T. R. (2011). Two-stage instrumental variable methods for estimating the causal odds ratio: analysis of bias.
\textit{Statistics in Medicine}, {\bf 30}, 1809-1824.
\medskip

\noindent Didelez, V. and Sheehan, N. (2007). Mendelian randomization as an instrumental variable approach to causal inference. \textit{Statistical Methods in Medical Research} {\bf 16}, 309-330.
 \medskip

\noindent Hern\'an, M. A. and Robins J. M. (2006). Instruments for causal inference: an epidemiologist's dream?
\textit{Epidemiology} {\bf 17}, 360-372.
\medskip

\noindent	Imbens, G. W. and Angrist, J.  (1994). Identification and estimation of local average treatment effects.
 \textit{Econometrica} \textbf{62}, 467-476.
 \medskip
 
 \noindent  Joffe, M.M., Yang, W.P. and Feldman, H. (2012). G-Estimation and Artificial Censoring: Problems, Challenges, and Applications. \textit{Biometrics} {\bf 68}, 275-286.
 \medskip
 
 \noindent  Kjaersgaard, M.I.S. and Parner, E. T.  (2016). Instrumental Variable Method for Time-to-Event Data Using aPseudo-Observation Approach. \textit{Biometrics} {\bf 72}, 463-472.
 \medskip

\noindent
Li, J.,  Fine, J.  and  Brookhart, A. (2014).  Instrumental variable additive hazards models.
\textit{Biometrics}, \textbf{71}, 122-130.
 \medskip

\noindent
 Martinussen, T. and Scheike, T. H. (2006). Dynamic Regression Models for Survival Data
Springer-Verlag New York
 \medskip

\noindent
Martinussen, T., Vansteelandt, S., Tchetgen Tchetgen, E. J. and Zucker, D. M. (2017). Instrumental variables estimation of exposure effects on a time-to-event endpoint using structural cumulative survival models. \textit{Biometrics}. doi:10.1111/biom.12699
\medskip

\noindent
Richardson, A., Hudgens, M. G.,   Fine, J.  and  Brookhart, A. (2017). 
Nonparametric binary instrumental variable analysis of competing risks data.
 \textit{Biostatistics} {\bf 18}, 48-61.
\medskip

\noindent
Robins, J.M. and Rotnitzky, A. (2004). Estimation of treatment effects in randomised trials with non-compliance and a dichotomous outcome using structural mean models. \textit{Biometrika} {\bf 91}, 763-783.
\medskip

\noindent
Robins, J.M. and Tsiatis, A. (1991). Correcting for non-compliance in randomized trials using rank-preserving structural failure time models. \textit{Communications in Statistics} {\bf 20}, 2609-2631.
\medskip

\noindent
Shapiro, S.  (1977). Evidence of screening for breast cancer from a randomised trial.
\textit{Cancer} {\bf 39}, 2772-2782.
\medskip

\noindent
Tchetgen Tchetgen, E. J., Walter, S., Vansteelandt, S., Martinussen, T., Glymour, M. (2015). Instrumental variable estimation in a survival context.
\textit{ Epidemiology} {\bf26}, 402-410.
\medskip

\noindent
Vansteelandt, S. and Goetghebeur, E. (2003). Causal inference with generalized structural mean models. \textit{Journal of the Royal Statistical Society, Series B} \textbf{65}, 817- 835.
\medskip

\noindent
Vansteelandt, S., Bowden, J., Babanezhad, M. and Goetghebeur, E. (2011). On instrumental variable estimation of the causal odds ratio.
\textit{Statistical Science}, \textbf{26}, 403-422.
\medskip

\noindent	 Zheng,  C. , Dai, R., Hari, P. N. and Zhang, MJ. (2017). 
Instrumental variable with competing risk model  \textit{Statistics in Medicine},
 doi: 10.1002/sim.7205.
 \medskip


\end{document}